\documentclass[aps,natphysics,superscriptaddress,two column,balance,notitlepage,bibnotes]{revtex4-2}
\usepackage{latexsym}
\usepackage{subcaption}
\usepackage{balance}
\usepackage{caption}
\usepackage[margin=0.7in]{geometry}
\usepackage[dvips]{color}
\usepackage[pdftex]{graphicx}
\usepackage{amsmath}
\usepackage{amssymb}
\usepackage{esdiff}
\usepackage{ragged2e}
\usepackage{lineno}
\usepackage{gensymb}
\usepackage{enumerate}
\usepackage{hyperref}
\usepackage{mathrsfs}
\usepackage[usenames,dvipsnames]{xcolor}
\usepackage{float}
\def\mr{\mathrm}

\begin{document}

\title{\textbf{Emergence of broken-symmetry states at half-integer band fillings in twisted bilayer graphene}}
\author{Saisab Bhowmik}
\email{saisabb@iisc.ac.in}
\affiliation{\textit{Department of Instrumentation and Applied Physics, Indian Institute of Science, Bangalore, 560012, India}}
\author{Bhaskar Ghawri}
%\email{xxx}
\affiliation{\textit{Department of Physics, Indian Institute of Science, Bangalore, 560012, India}}
\author{Nicolas Leconte}
\affiliation{\textit{Department of Physics, University of Seoul, Seoul 02504, Korea}}
\author{Samudrala Appalakondaiah}
\affiliation{\textit{Department of Physics, University of Seoul, Seoul 02504, Korea}}
\author{Mrityunjay Pandey}
%\email{xxx}
\affiliation{\textit{Centre for Nano Science and Engineering, Indian Institute of Science, Bangalore 560 012, India}}
\author{Phanibhusan S. Mahapatra}
%\email{xxx}
\affiliation{\textit{Department of Physics, Indian Institute of Science, Bangalore, 560012, India}}
\author{Dongkyu Lee}
\affiliation{\textit{Department of Physics, University of Seoul, Seoul 02504, Korea}}
\affiliation{\textit{Department of Smart Cities, University of Seoul, Seoul 02504, Korea}}
\author{K. Watanabe}
%\email{xxx}
\affiliation{\textit{Research Center for Functional Materials, National Institute for Materials Science, Namiki 1-1, Tsukuba, Ibaraki 305-0044, Japan}}
\author{ T. Taniguchi}
%\email{xxx}
\affiliation{\textit{International Center for Materials Nanoarchitectonics, National Institute for Materials Science, Namiki 1-1, Tsukuba, Ibaraki 305-0044, Japan}}
\author{Jeil Jung}
\email{jeiljung@uos.ac.kr}
\affiliation{\textit{Department of Physics, University of Seoul, Seoul 02504, Korea}}
\affiliation{\textit{Department of Smart Cities, University of Seoul, Seoul 02504, Korea}}
\author{Arindam Ghosh}
\affiliation{\textit{Department of Physics, Indian Institute of Science, Bangalore, 560012, India}}
\affiliation{\textit{Centre for Nano Science and Engineering, Indian Institute of Science, Bangalore 560 012, India}}
\author{U. Chandni}
\email{chandniu@iisc.ac.in}
\affiliation{\textit{Department of Instrumentation and Applied Physics, Indian Institute of Science, Bangalore, 560012, India}}

%\date{07/06/2021}
\maketitle
\textbf{
The dominance of Coulomb interactions over kinetic energy of electrons in narrow, non-trivial moir\'{e} bands of magic-angle twisted bilayer graphene (TBG) gives rise to a variety of correlated phases such as correlated insulators \cite{cao2018correlated,lu2019superconductors,PhysRevX.8.031089}, superconductivity \cite{cao2018unconventional,lu2019superconductors,yankowitz2019tuning,yankowitz2019tuning,PhysRevLett.121.257001,PhysRevLett.122.257002}, orbital ferromagnetism \cite{lu2019superconductors,sharpe2019emergent,serlin2020intrinsic}, Chern insulators  \cite{nuckolls2020strongly,wu2021chern,saito2021hofstadter,das2021symmetry} and nematicity \cite{cao2021nematicity}. Most of these phases occur at or near an integer number of carriers per moir\'{e} unit cell.~Experimental demonstration of ordered states at fractional moir\'{e} band-fillings at zero applied magnetic field $B$, is a challenging pursuit.~In this letter, we report the observation of states near half-integer band-fillings of $\nu\approx 0.5$ and $\pm3.5$ at $B\approx 0$ in TBG proximitized by tungsten diselenide (WSe$_2$) through magnetotransport and thermoelectricity measurements. A series of Lifshitz transitions due to the changes in the topology of the Fermi surface implies the evolution of van Hove singularities (VHSs) of the diverging density of states (DOS) at a discrete set of partial fillings of flat bands. Furthermore, at a band  filling of $\nu\approx-0.5$, a symmetry-broken Chern insulator emerges at high $B$, compatible with the band structure calculations within a translational symmetry-broken supercell with twice the area of the original TBG moir\'{e} cell.~Our results are consistent with a spin/charge density wave ground state in TBG in the zero $B$-field limit.\\
}

\begin{figure*}[bth]
\includegraphics[width=1.0\textwidth]{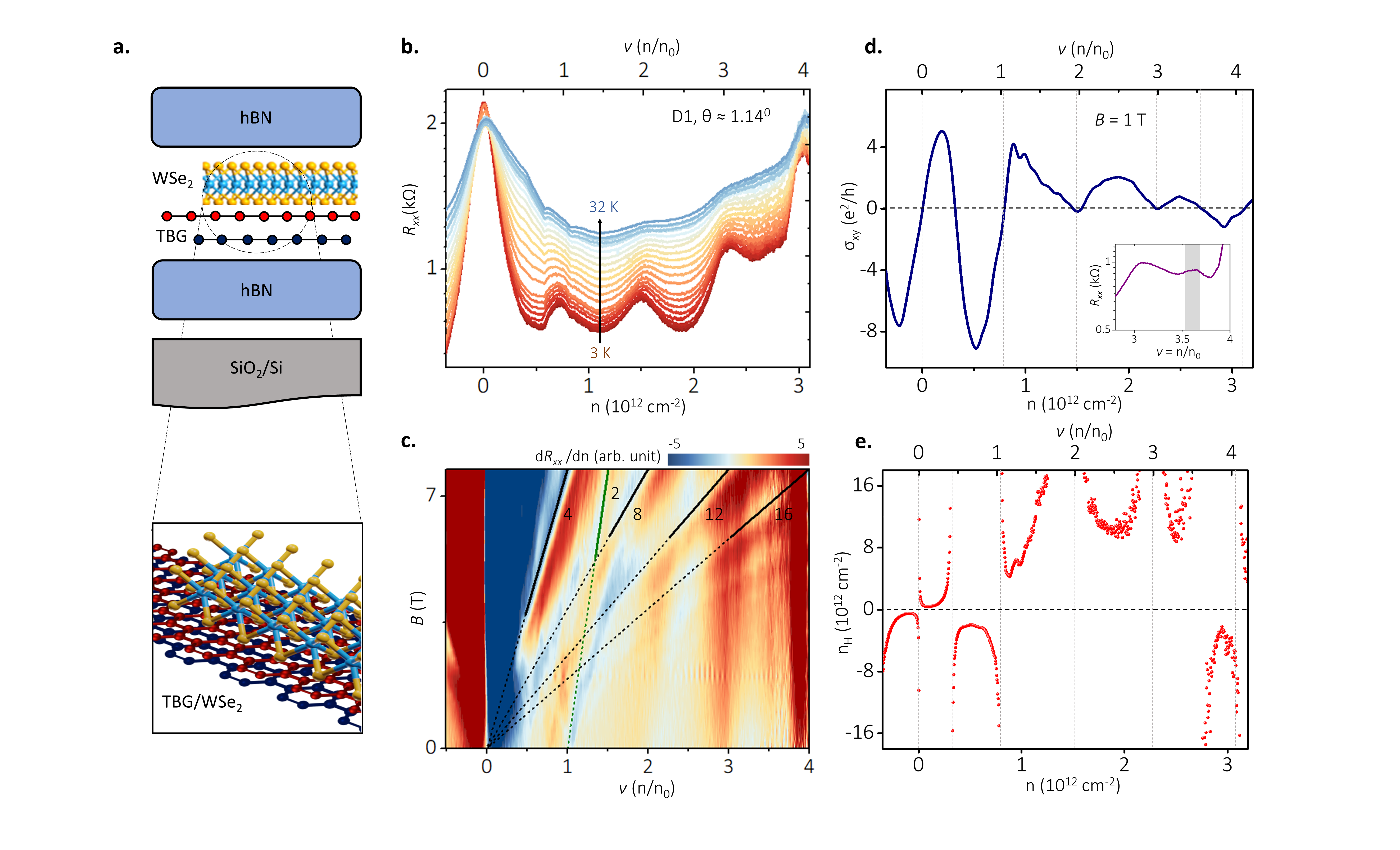}
\captionsetup{justification=raggedright,singlelinecheck=false}
\justify{FIG. 1. \textbf{Electrical characterization and low-field Hall measurements in device D1, $\theta \approx 1.14^\circ$}. \textbf{a.} Schematic of hBN encapsulated TBG/WSe$_2$ heterostructure on SiO$_2$/Si substrate. \textbf{b.} Four-probe longitudinal resistance $R_{xx}$ as a function of carrier density $n$ measured in the temperature $T$ range $3 - 32$~K for $B = 0$. The twist angle $\theta$ $\approx$ $1.14^{\circ}$  was calculated from the position of the resistance peak at a filling $\nu$ ($n$/$n_0$) = 4 where $n_0$ is the density corresponding to one carrier per moir\'{e} unit cell. \textbf{c.} Landau fan diagram at $T = 5$ K for magnetic field $B$ up to 7.75 T where d$R_{xx}$/d$n$ is plotted as a function of ($\nu$, $B$). The Landau levels (LL) diverging from the charge neutrality point (CNP) are shown by solid black lines. The filling factors 4, 8, 12, and 16 are determined from the slopes of these lines. In addition, a straight line with a slope of 2 that nucleates from $\nu$ = 1 at B $\gtrsim$ 5~T is shown by the solid green line. \textbf{d.} Density dependence of field-symmetrized Hall conductivity $\sigma_{xy}$ at $B=1$ T and $T$ = 6 K. Zero-crossings appear at $\nu$ = 0, 0.5, 1, 3.5 and 4. $\sigma_{xy}$ suddenly drops to zero at $\nu$ = 2 and 3 and rises again upon crossing $\nu$ = 2 and 3. The inset shows an additional peak in $R_{xx}(n)$ around $\nu$ = 3.5, when $B=0$. This data in $R_{xx}$ was recorded after a magnetic field sweep and the peak at $\nu$ = 3.5 persisted at zero-field for the next few thermal cycles. \textbf{e.} Hall density $n_H$, calculated from Hall resistance $R_{xy}$ exhibits similar behavior as $\sigma_{xy}$.}
\end{figure*}

\begin{figure*}[bth]
\centering 
\includegraphics[width=1.0\textwidth]{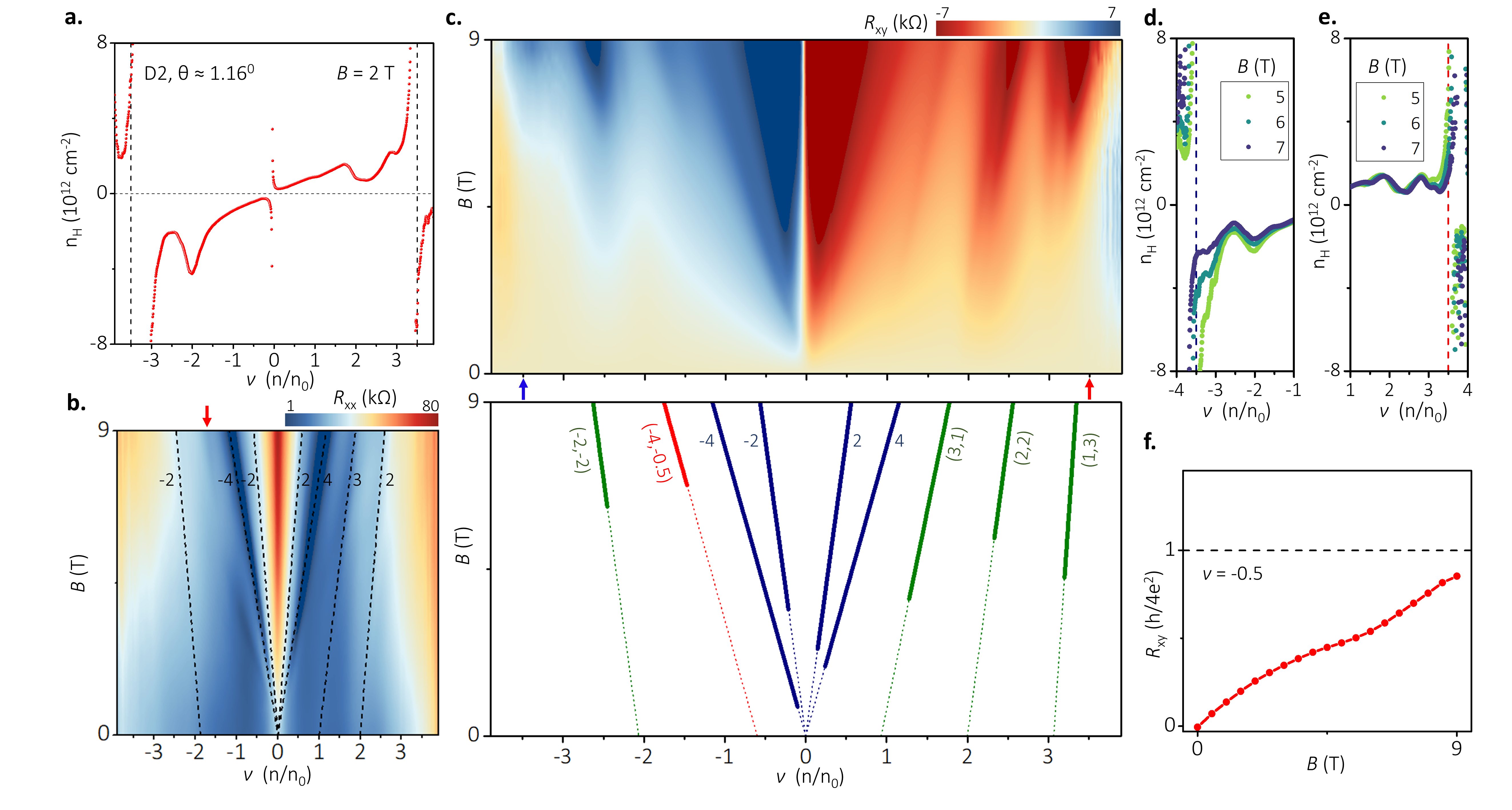}
%\captionsetup{justification=raggedright,singlelinecheck=false}
\justify{FIG. 2. \textbf{Symmetry-broken states at $\nu \approx -0.5$ and $\pm3.5$ in device D2, $\theta \approx 1.16^\circ$}. \textbf{a.} $n_H$ vs. $\nu$ at $B = 2$ T and $T = 5$ K shows sign-changes at $\nu = 0, \pm3.5$ and reset at $\nu = \pm 2$. \textbf{b.} $R_{xx}$ as a function of $\nu$ for different $B$ up to 9 T and the color bar shown here is in logarithmic scale.~The minima in $R_{xx}$ emanating from $\nu = 0$ have slopes of $\pm2, \pm4$, while those from $\nu = 1$ and $\pm2$ have slopes $3$ and $\pm2$, respectively. An additional state develops at $B\gtrsim6.5$ T whose position in density at $B = 9$ T is marked by the red arrow on the top axis. \textbf{c.} $R_{xy}$ vs ($\nu,B$) at $T = 5$~K shows multiple states at different $\nu$. The red and blue arrows in the bottom axis mark $\nu = +3.5$ and $\nu = -3.5$ respectively, where $R_{xy}$ changes sign. In the bottom panel, these states are labelled by ($C,\nu$) = ($\pm4,0), (\pm2,0)$ (solid blue), ($1,3), (\pm2,2), (3,1)$ (solid green) and ($-4,-0.5$) (solid red). \textbf{d-e.} $n_H$ exhibits sign changes at $\nu = \mp3.5$ for high magnetic fields (5-7 T). \textbf{f.} $R_{xy}$ at $\nu \approx -0.5$ for different $B$, along the red line in the lower panel in (c) that approaches the quantized limit $R_{xy} = h/4e^2$ %represented with the horizontal dashed black line 
as the field is increased.}
\end{figure*}

\begin{figure}[bth]
\centering 
\includegraphics[width=1.0\columnwidth]{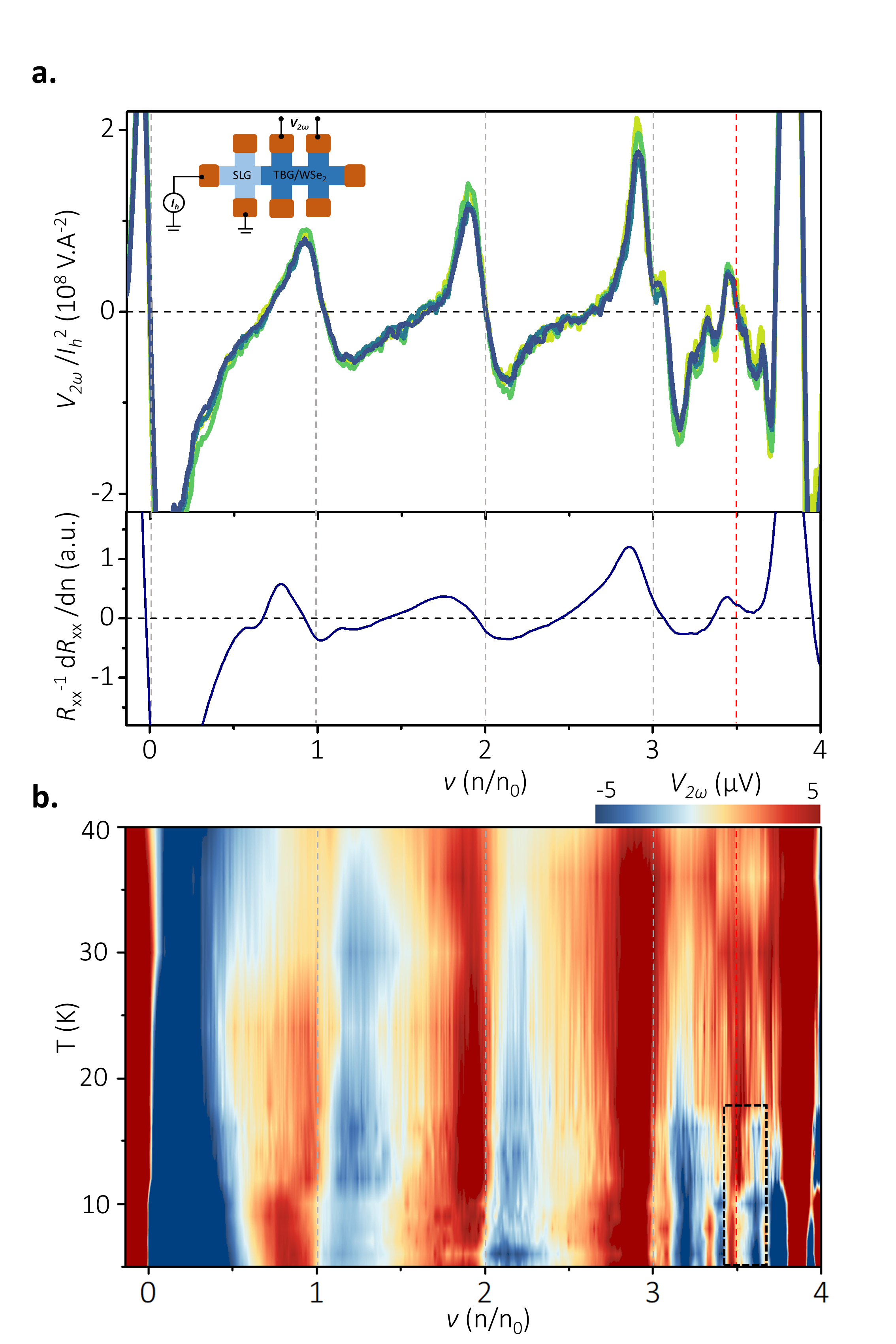}
%\captionsetup{justification=raggedright,singlelinecheck=false}
\justify{FIG. 3. \textbf{Density-dependence of zero-field thermoelectricity}. \textbf{a.} Second harmonic component of thermo-voltage $V_{2\omega}$ as a function of $n$ for heating currents $I_h = 300-400$ nA, with a step size of 20 nA, measured at $T = 5$ K. A schematic of the heating geometry and measurement of $V_{2\omega}$ can be found in the inset. For all $I_h$, the values of $V_{2\omega}/I_h^2$ are constant, which indicates the linear response regime in the range of $I_h$ used for heating. Sign-reversals at $\nu$ = 1, 2, 3 are marked by vertical dotted lines (grey). An additional zero-crossing appears at $\nu$ = 3.5 (red dotted line). ($1/R_{xx})dR_{xx}/dn$ is shown in the bottom panel.~Both $V_{2\omega}$ and $R_{xx}$ are measured independently in the same thermal cycle. \textbf{c.} Temperature dependence of $V_{2\omega}$ in the range $5 - 40$~K. We observe a series of sign changes as $n$ is swept across the different correlated phases. Two vertical regions with opposite signs in $V_{2\omega}$ on the low and high density sides of $\nu$ = 3.5 are outlined by a rectangular window for clarity.~The zero-crossing at $\nu$ = 3.5 disappears around $T$ $\gtrsim$ 18~K.}
\end{figure}

%\section*{Main}
In two-dimensional electron systems, a quantizing magnetic field $B$ is an essential ingredient for the experimental realization of both integer and fractional quantum Hall states. The first proposal for a zero-$B$ analogue of the integer quantum Hall effect (QHE) came from the Haldane model \cite{PhysRevLett.61.2015}, where non-interacting electrons hopping on a honeycomb lattice with a zero-average, spatially inhomogeneous $B$ exhibit quantized Hall conductances.~Subsequently, interacting lattice models with nearly flat Chern bands proposed fractional QHE in the absence of $B$~\cite{SDSarmaPRL,PhysRevLett.106.236804}. The key to realizing these fractional states is the search for materials with extremely flat bands featuring non-trivial topological properties.~TBG, where two graphene sheets are stacked together with a rotational mismatch near the first magic-angle of $1.1^{\circ}$, is an ideal material platform to investigate such zero-$B$ states.~Recent experiments have revealed a plethora of broken-symmetry phases in TBG \cite{cao2018correlated,lu2019superconductors,cao2018unconventional,lu2019superconductors,yankowitz2019tuning,stepanov2020untying,sharpe2019emergent,serlin2020intrinsic,nuckolls2020strongly,wu2021chern,saito2021hofstadter,das2021symmetry,cao2021nematicity,zondiner2020cascade} and related moir\'{e} materials \cite{polshyn2021topological,Hao1133,liu2020tunable,rickhaus2020densitywave}, where the external tuning of several parameters leads to the desired symmetry breaking. For instance, $C\textsubscript{2}$ inversion symmetry  can be broken by aligning the TBG layers with hexagonal Boron Nitride (hBN) \cite{serlin2020intrinsic}, while the application of a %large 
$B$-field breaks time reversal symmetry $T$.~These experimental knobs enable access to the low energy degrees of freedom of the band structure, which hosts non-trivial topological, as well as trivial band insulators at or near integer fillings of the moir\'{e} unit cell.~Incompressible states stabilized by electron-electron interactions within the flat band, were found to appear at large $B$-fields, in the form of Hall plateaus featuring Chern numbers $C$ at various band filling index $s$ \cite{nuckolls2020strongly,wu2021chern,saito2021hofstadter,das2021symmetry,PhysRevLett.123.036401}.~Recently, transport measurements in TBG uncovered a high $B$-field-induced state at $s$ = 3.5, suggesting the possibility of a new VHS and fractional Chern insulator \cite{wu2021chern}.~However, states originating from fractional fillings of the unit cell at zero or weak $B$-field have remained elusive in TBG devices, although a recent report on twisted monolayer-bilayer graphene indicates incompressible insulators with $C=1$ at fractional band fillings of $s=$ 1.5 and 3.5 \cite{polshyn2021topological}. Notably, states at several fractional fillings have been observed in transition metal dichalcogenide (TMD)-based moir\'{e} systems, attributed to generalized Wigner crystallization of the underlying lattice and charge density waves ordering \cite{xu2020correlated}.~States featuring fractional $C$ and/or fractional $s$, which unify the possibility of many exotic states such as spin/charge density wave phases (fractional $s$), symmetry broken Chern insulators (integer $C$, fractional $s$), and fractional Chern insulators (fractional $C$ and $s$), are thus of great interest to our growing understanding of this fascinating material platform. While all these states emerge from strong electron-electron interactions within the flat bands, subtle changes in experimental configurations can modify the anisotropy energy that decides the preferred ordered phase when the degenerate phases are in close competition~\cite{PhysRevB.103.125406,PhysRevLett.123.096802}.\\

In this work, we report the observation of novel states at half-integer moir\'{e} band fillings in TBG/WSe$_2$ devices, that persist at $B\approx 0$, suggesting the possibility of a spin/charge density wave order.~We focus on magnetotransport and thermoelectric transport measurements on two TBG/WSe$_2$ moir\'{e} heterostructures D1 and D2, with twist angles $\theta\approx1.14^{\circ}$ and $1.16^{\circ}$ respectively, fabricated using the `tear and stack' technique (see Methods and Supplementary Fig.~S6)~\cite{kim2016van,cao2018unconventional}.~Few-layer WSe$_2$ flake of thickness $\approx$ 2-3 nm was placed on the TBG, sandwiched between two hBN layers, with the stack assembled on SiO$_2$/Si substrate (Fig.~1a).~Characterization for D1 starts in Fig.~1b
with the four-probe longitudinal resistance $R_{xx}$ as a function of charge carrier density ($n$)
in the temperature ($T$) range of 3$-$32~K for $B=0$~T. The hole side in device D1 was inaccessible due to dielectric leakage. We observe correlated states at all partial fillings $\nu$ = 1, 2, and 3 and a resistive peak at $\nu$ = 4, where the conduction band is fully occupied. The twist angle ($\theta \approx$ $1.14^{\circ}$) is estimated from the density at the superlattice gap ($\nu=4$) (see Methods).~The resistance at $\nu$ = 1, 2, 3 increases with increasing temperature in the measured range, and the peaks disappear at higher $T$. Further, we observe that linear-in-$T$ behavior of $R_{xx}(n)$ at $\nu$ = 1, 2 and 3 persists up to $T = 32$~K (see Supplementary Fig.~S8).~Device D2 also exhibits similar correlated states at different $\nu$ across various contact configurations (see Supplementary Fig.~S7).~Fig.~1c shows the Landau fan diagram for device D1 in a perpendicular $B$-field, where the linearly dispersing features emanating from the charge neutrality point (CNP) and other band fillings can be characterized by fitting the Diophantine equation, $n/n_0 = C\phi/\phi_0 + s$, where $n_0$ is the density corresponding to one carrier per moir\'{e} unit cell, $\phi$ is the magnetic flux per moir\'{e} unit cell, $\phi_0 = h/e$ is the flux-quantum, $h$ is Planck's constant, $e$ is the charge of electron, and $s$ is the band filling index or the number of carriers per unit cell at $B = 0$ T. The slope gives the Chern number via the Streda formula, $C$ ($=h/e$ $\partial n/\partial B$).~The Landau-levels (LLs) diverging from the CNP have filling factors of 4, 8, 12, and 16 (black lines) that increase in steps of 4 due to the four-fold spin/valley degeneracy. For $B\gtrsim$ 5~T, we observe an additional state that can be extrapolated to $\nu$ = 1 at $B=0$, with a slope of 2 (green line). We now focus on the low $B$-field Hall measurements. Fig.~1d shows Hall conductivity $\sigma_{xy}$($n, B$) as a function of $n$ at a perpendicular $B$ = 1 T. We find that, while zero-crossings of $\sigma_{xy}$ are observed, as expected, at the CNP and the band insulator at $\nu=$ 4, a `reset' behaviour is observed at $\nu$ = 2, 3, namely 
% Near close vicinity of $\nu$ = 2 (3), 
$\sigma_{xy}$ decreases to zero 
% at $\nu$ = 2 (3) 
and rises again through $\nu$ = 2 and 3 without any sign change. These features have been reported in earlier studies of TBG and related systems~\cite{wu2021chern,saito2020independent}, and have been attributed to enhanced Coulomb interactions within the malleable bands that favor a gap opening by splitting the band into a low-energy filled and high-energy empty sub-band when the Fermi energy $E_F$ crosses the VHS at $\nu=2$. A similar splitting of the bands that occur at the consequently developed VHS at $\nu=3$, leads to yet another reset~\cite{wu2021chern}. 
Surprisingly, 
% and differently to the aforementioned earlier work, %  
some novel observations with respect to earlier work are the
additional zero-crossings in $\sigma_{xy}(n)$ at $\nu=1$ and half-integer $\nu=0.5$ and 3.5. Intriguingly, a minor peak in $R_{xx}(n)$ around $\nu$ = 3.5 was seen even at $B = 0$ in several later thermal cycles as shown in the inset to Fig.~1d. It is often illustrative to plot the Hall density $n_H$, calculated from Hall resistance $R_{xy}$ as $n_H$ = ${-\frac{1}{e}}{\frac{B}{R_{xy}}}$, which provides insights into the VHSs and Fermi surface topology. In Fig.~1e, $n_H$ exhibits features consistent with $\sigma_{xy}$, diverging with opposite signs on both low and high-density sides of $\nu$ = 0.5, 1 and 3.5. Furthermore, no sign-reversal in $n_H$ is observed at $\nu$ = 2, 3. Remarkably, we find robust states at half-integer $\nu$ even at $B$-fields as low as $\sim$0.3~T in device D1 (see Supplementary Fig.~S9).\\

To further investigate the presence and robustness of states at fractional fillings we report in Fig.~2 detailed magnetotransport measurements in a second device D2. We observe similarities and differences, the latter possibly due to specific device details.~Consistent with device D1, sign changes are observed in $n_H$ at $\nu = \pm3.5$, down to $B = 0.2$ T (see Fig.~2a for $B = 2$~T and supplementary Fig.~S10), which remain robust at high $B$ as well (Fig.~2d-e). Further, $n_H$ shows reset of charge carriers at $|\nu| = 2$ similar to device D1. Fig.~2b and 2c show the Landau fan diagrams in $R_{xx}$ and $R_{xy}$ respectively. The filling factors of $\pm2$ emanating from the CNP indicate that the four-fold spin/valley symmetry is broken.~Additionally, multiple linearly dispersing minima in $R_{xx}$ are observed at high $B$, originating from various $s$.~We also note that these minima in $R_{xx}$ are accompanied by the wedge-like resistive features in $R_{xy}$ (Fig.~2c).~Our observation of possible Chern insulators, labelled by the two indices $C$ and integer $s$ with ($C$, $s$) = ($\pm2, \pm2$), ($1, 3$) and ($3,1$) are in a good agreement with recent reports on TBG systems~\cite{nuckolls2020strongly,wu2021chern,saito2021hofstadter,das2021symmetry}.~Chern insulators at integer $s$ can be understood within the two spin/valley degenerate flat band model in TBG where $C_2T$ symmetry breaking opens a gap at the CNP and produces isolated flat Chern bands with opposite Chern numbers $C=+1$ and $C=-1$~\cite{PhysRevB.99.195455,saito2021hofstadter,wu2021chern,nuckolls2020strongly,das2021symmetry}. 
The most striking feature in our data is a new state at half-integer $s \approx -0.5$ for $B\gtrsim6.5$~T (denoted by red arrow in Fig.~2b and red line in the lower panel of Fig.~2c) for which we obtain $C=-4$ from Streda formula in both $R_{xx}$ and $R_{xy}$. The magnetic field evolution of the state along the same red line in Fig.~2c is shown in Fig.~2f, where we observe that $R_{xy}$ approaches $h/4e^2$ within the experimental range of $B$-field, providing further evidence for the Chern insulating state.\\ 

\begin{figure}[h]
\centering 
\includegraphics[width=0.8\columnwidth]{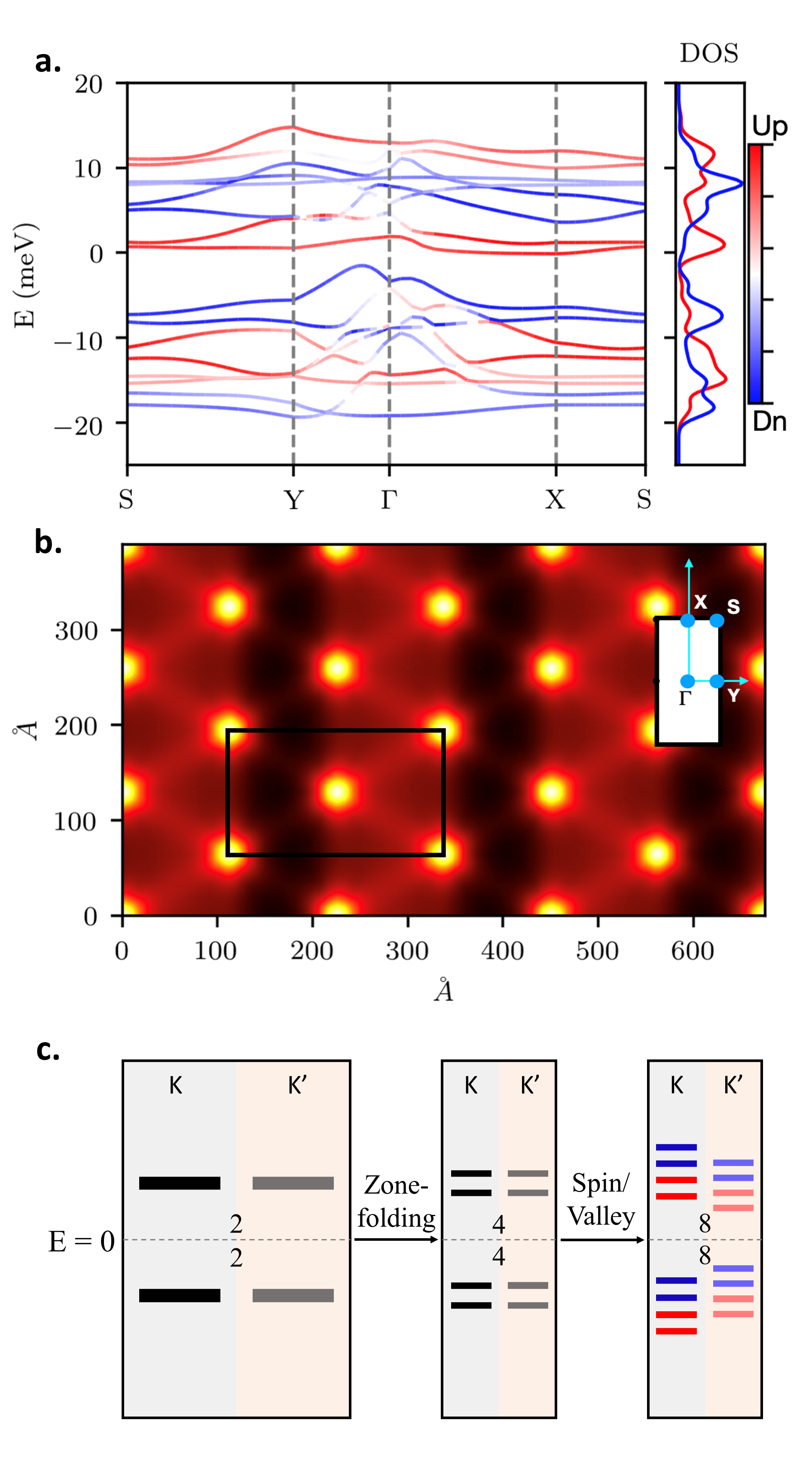}
%\captionsetup{justification=raggedright,singlelinecheck=false}
\justify{FIG. 4. \textbf{Degeneracy lifting of the folded bands}. 
\textbf{a.} Spin-resolved band structure and corresponding DOS of TBG under a 1D periodic potential given in Eq.~\ref{Heq} with $C_0 = 0$, $C_z = 30$~meV, $\phi = 0$, 
$\Delta = 300$~meV,  $t_2 = 15$~meV and $\lambda_{\rm R} = 10.5$~meV.
The red and blue shades indicate the strength of the respective spin orientations induced by the SOC terms.
\textbf{b.} Schematic that combines the colormap of the interlayer distances from the relaxed TBG system (AA stackings appear as bright dots) together with the 1D sine potential whose zero is at AA-stacking 
to illustrate the rectangular cell whose area is twice as large as that of TBG moir\'{e} unit cell. 
The inset represents the high symmetry points and the $k$-path corresponding to the band structure in the main panel. \textbf{c.} The sketch illustrates how the zone-folding doubles the number of flat bands and halves the Brillouin zone area, while splitting of the spin and valley degrees of freedom bring the total number of bands to 16. The high/low intensity of colors for the lines distinguish the $K$ and $K'$ valleys and the red and blue colors refer to the spin degree of freedom. The valley-degeneracy lifting is illustrated by a small shift in the K$^\prime$ valley levels with respect to the bands in the K valley.}
\end{figure}

While Hall measurements demonstrate the existence of states at fractional fillings for low $B$, the technique precludes direct observation of these states at $B=0$. With this aim, we have performed thermoelectric transport measurements at $B=0$ for device D1. The experimental technique used for measurement of the thermo-voltage is briefly discussed below and in methods section. A temperature gradient ($\Delta T$) over the entire TBG/WSe$_2$ channel is created by passing a sinusoidal current ($I_h$) through SLG contacts outside the twisted region, shown in the inset of Fig.~3a. We measure the second-harmonic component of thermo-voltage $V_{2\omega}$ for different $I_h$ (described in detail elsewhere \cite{ghawri2020excess}) in a density range from the CNP to $\nu$ = 4 (Fig.~3a). The linear variation of $V_{2\omega}$ with ${I_h}^2$ ensures a linear response regime for the range of heating currents used. The $n$-dependence of $V_{2\omega}$ shows a series of zero-crossings at $\nu$ = 0, 1, 2, 3. Most importantly, we observe that the sign-reversal of $V_{2\omega}$ also occurs at half-integer $\nu$ = 3.5 at $B = 0$. In Fermionic systems at very low temperatures ( $T\ll T\textsubscript{F}$, where $T_F$ is the Fermi temperature), where scattering of charge carriers is diffusive, semiclassical Mott relation (SMR) establishes a connection between thermoelectric power ($S=V_{2\omega}/\Delta T$) and electrical resistance ($R_{xx}$) \cite{PhysRevB.21.4223,ghawri2020excess} as,
\begin{equation}
S_\mr{Mott}
= \frac{\pi^2 k_\mr{B}^2 T}{3|e|} \frac{1}{R_{xx}} \frac{\mr{d}R_{xx}}{\mr{d}n} \frac{\mr{d}n}{\mr{d}E}\bigg\vert_{E_\mr{F}}, 
\label{Motteq}
\end{equation}
where $\frac{\mr{d}n}{\mr{d}E}$ represents the DOS.
In deriving the SMR, a non-interacting scenario is assumed with the electrical conductance varying smoothly near $E_{F}$. SMR holds true in a large number of van der Waals materials, including monolayer and bilayer graphene \cite{PhysRevLett.102.096807, jayaraman2021evidence} and disordered metals/insulators.~We compare $V_{2\omega}$ with derivative of $R_{xx}$ with respect to $n$, shown in the bottom panel of Fig.~3a. We observe exact mapping of zero-crossings at $\nu = 1, 2 $ and 3, with peaks in $R_{xx}$. While, $\frac{dR_{xx}}{dn}$ shows a weak peak at $\nu=3.5$, a clear sign change is observed in $V_{2\omega}$.~Intriguingly, the state at $\nu=0.5$ does not appear at $B=0$ in either $R_{xx}$ or $V_{2\omega}$. This we believe is because the sign changes in $V_{2\omega}$ occur in accordance with the resistive peaks in $R_{xx}$, while the sign changes in $\sigma_{xy}$ and $n_H$ are intricately connected to the Fermi surface topology via $R_{xy}$.~The temperature dependence of $V_{2\omega}$ is illustrated in Fig.~3b.~The sign reversals are seen as a series of vertical red and blue bands throughout the measured $T$ range. While the transitions at the CNP and $\nu$ = 1, 2 are robust and persist up to the measured temperature of $T = 40$ K, the feature at $\nu$ = 3.5 vanishes for $T \gtrsim$ 18~K.\\

The emergence of distinct features in magnetotransport and thermoelectricity at half-integer band fillings for two different devices, despite the inevitable variability typically seen in TBG samples, suggest a robust mechanism behind our observations.~We speculate that strong correlations between electrons may generate a spin/charge density order that breaks translation symmetry in the moir\'{e} leading to Brillouin zone (BZ) folding with electron wave functions extending over a unit cell with doubled area. In the non-interacting, single-particle band structure of TBG, a VHS is expected to appear at $|\nu|=2$, which is the half-filling of the first low-energy non-dispersive conduction or valence band, separating a two-pocket Fermi surface around the $K-$points from a higher-energy single-pocket Fermi surface around the $\Gamma$-point of the moir\'{e} BZ. This has been consistently verified in transport experiments~\cite{kim2016charge,wu2021chern,cao2018unconventional}. Broken translational symmetry and appearance of folded BZ in the doubled moir\'{e} unit cell may favor VHS at $\nu = 1$ and other partial fillings instead, as observed in our devices.~While higher order translational symmetry breaking (tripling, quadrupling) is a possibility, we can not rule out the presence of an inherent, accidental VHS at $\nu = 0.5$ in the TBG/WSe$_2$ band structure either. As noted in the theoretical model below, additional spin splitting of bands can lead to a complex DOS hierarchy, which can explain the VHS features at half integer fillings.~Furthermore, the emergence of a Chern insulator at $\nu\approx-0.5$ in device D2, strengthens the possibility of a doubled moir\'{e} unit cell. We also remark that the observation of a state with a slope of $C=2$ as obtained from Streda formula from a band filling of $s=1$ for device D1, is reminiscent of an odd parity ($p=C+s$) Chern insulator, recently reported in compressibility measurements of TBG due to reconstruction of Chern bands driven by the modulation of charge density waves over the real space moir\'{e} unit cell~ \cite{pierce2021unconventional}.\\

In the following we illustrate in Fig.~4a the band structure of TBG subject  to a 1D density wave potential 
that doubles the moir\'{e} unit cell area (Fig. 4b) and hence the number of bands,
similar to the proposal in twisted monolayer-bilayer graphene~\cite{polshyn2021topological}. 
The four-fold degenerate conduction and valence bands would turn into two eight-fold degenerate bands
upon zone-folding when the supercell area is doubled.~In Fig.~4c, we schematically illustrate the degeneracy lifting of these bands resulting in a total of sixteen split bands
in the presence of spin/charge density potential patterns. 
The states at half-integer fillings observed in experiments
are therefore compatible with an odd integer filling of the zone-folded split bands.
Our proposal of 1D patterned phases whose periods are commensurate with the moir\'{e} patterns can give rise to fractionally filled flat bands, analogous to the Bloch band filling index $s = \pm 1/3, 1/2$ states seen in Landau levels~\cite{Wang1231} that can be justified in the presence of Wigner crystal phases~\cite{macdonald}.
Our tight-binding Hamiltonian in the carbon $\pi$-orbitals basis incorporates this 1D pattern through 
a sinusoidal function and further incorporates the sublattice staggering as well as
the intrinsic and Rashba spin-orbit coupling potentials
that are often introduced when describing proximity spin-orbit coupling of graphene on transition metal dichalcogenide (TMD).

\begin{equation}
\begin{split}
H = \sum_{l,m,,\sigma} t_{lm} c^\dagger_{l \sigma} c_{m \sigma} 
+  \sum_{l,\sigma} (C_0 + \xi_{l}  C_z) \sin(\frac{\pi x_l}{\lambda_{M}^{\rm TBG}} + \phi)  \, c^\dagger_{l \sigma} c_{l \sigma} \\
 + 
 \sum_{l_t,\sigma}  \xi_{l_t} \Delta c^\dagger_{l_t \sigma} c_{l_t \sigma} 
 + \sum_{\langle \langle l_t,m_t\rangle \rangle \sigma} \nu_{l_t m_t} t_2 e^{i \pi/2} c^\dagger_{l_t \sigma} c_{m_t \sigma} 
\\ 
+  \frac{2i}{3} \sum_{\langle l_t, m_t \rangle} \sum_{\sigma,\sigma^\prime} c^\dagger_{l_t \sigma} c_{m_t \sigma^\prime} \left[ \lambda_R (\hat{\textbf{s}}\times\textbf{d}_{l_t m_t})_z\right]_{\sigma \sigma^\prime}.
\label{Heq}
\end{split}
\end{equation}
Here, the first term describes the carbon interatomic hopping terms that are tuned to generate the magic angle flat bands at $1.08^\circ$~\cite{1910.12805}, see the supplementary information for the expression of the distance-dependent $t_{lm}$ terms. The  $l, \, m$ indices are lattice labels $l_t, \, m_t$ with the $t$ subscript referring to top layer sites that are contacting the TMD. The second 1D sine term describes a density wave potential repeating with a period of $2\lambda_{M}^{\rm TBG}$, twice the TBG moir\'{e} period,which defines a supercell that doubles the moir\'{e} area and allows to obtain folded bands that split at half-integers of the unperturbed system's filling factors. The $\phi$ phase term is a control knob that shifts the potential origin position. The remaining three terms capture the WSe$_2$ interface effect in the contacting graphene layer through a constant mass term with unequal sublattice resolved potentials, and through Kane-Mele and Rashba SOC models, where $\nu_{l_t m_t}=\pm 1$ following the conventions in Ref.~\cite{PhysRevLett.106.236804} where $\hat{\textbf{s}}$ are the spin Pauli matrices and $\textbf{d}_{l_t m_t}$ is the displacement vector between lattice sites $l_t$ and $m_t$. These terms introduce an energy anisotropy that favors a certain spin/valley isospin polarization of the ground-state~\cite{fabian}.~Because the lattice constant mismatch between TBG and WSe$_2$ is already large, the proximity spin-orbit coupling terms~\cite{Koshino_tmd} in graphene do not change drastically when we modify the twist angle from alignment. The system parameters for the band structures are {\em ad hoc} choices that illustrate the degeneracy lifting of the folded flat bands compatible with the observed half-integer filling of $\nu = 0.5$ and $3.5$, and the gaps at the integer filling of $\nu = 2$ and $3$ (partially). The simultaneous presence of mass with opposite $\xi_{l} = \pm 1$ signs for $A, \, B$ sublattice resolved potentials and SOC terms can cause valley and spin splitting that lifts the degeneracy between the flat bands which the Coulomb interactions can further amplify. We refer to the supplementary information for the bands associated with alternative $\phi= \pi/2$ cosine-like 1D potential that illustrates the overall robustness to shifts in its origin.\\

Our work shows that generation of flat bands by means of moir\'{e} superlattices is enriched by the possibility of superposing periodic potentials that are commensurate with the original moir\'{e} pattern~\cite{PhysRevB.103.075423}.~The favored ground state among several quasi-degenerate phases will depend on the anisotropy energy for the different spin/valley isospin polarization that can change with details in sample preparation and device assembly methods. Here, we have only considered the $s = 1/2$ Bloch band filling index scenario, while $s = \pm 1/3$ for Kekule patterns with three times larger superlattice area seen in experiments with Graphene/hBN~\cite{Wang1231,macdonald} also deserve attention in future research. Different specific material choices for the TBG dielectric environment are expected to subtly influence the ground state configuration. In particular, the way in which the spin and valley isospins order for various fillings will impact system properties like the spin or orbital magnetism stemming from the topological nature of the bands, the superconducting pairing symmetry as well as the instabilities in the electron-electron and electron-hole channels. Our work suggests that using TMD dielectrics is a promising pathway to further understand and explore the nature of fractional-filling correlated states in the limit of vanishing magnetic fields.~This prospect is ever brighter, thanks to the large variety of TMD materials that could be used to engineer moir\'{e} heterostructures.
\\

{\em Note added in proof} : While preparing our manuscript, we became aware of related observations of fractional Chern insulators in hBN-interfaced magic-angle TBG reported in \cite{xie2021fractional} at filling fractions different than in our work. \\

\section*{METHODS}

\subsection*{Device fabrication}

The devices in this work were fabricated using a modified `tear and stack' method (see Supplementary Fig.~S6).~A hBN flake ($10-22$~nm) was picked up using a polypropylene carbonate (PPC) film that was coated on a polydimethylsiloxane (PDMS) stamp. The hBN flake was then used to pick up a multilayer WSe$_2$ ($\sim 2-3$ nm). Next, hBN/WSe$_2$ was slowly placed on a monolayer graphene flake.~After picking up the first half of the graphene, torn by hBN, the transfer stage was rotated approximately by $1.1^{\circ}$, and the second half of the graphene was picked up. In the final step, a thick hBN flake ($25-30$~nm) was picked up, and the encapsulated stack was released on a 285 nm SiO$_2$/Si wafer at a temperature of $85^{\circ}$C. We set the temperature at $63^{\circ}$C for picking up hBN and $55^{\circ}$C for picking up graphene and WSe$_2$. The PPC film was then cleaned with anisole and organic solvents. We defined the Hall bar geometry of the final device using electron-beam lithography, reactive ion etching using CHF$_3$/O$_2$, and thermal evaporation of ohmic edge contacts using Cr/Au (5 nm/60 nm). WSe$_2$ crystals used in this work were obtained from 2D Semiconductors.\\

\subsection*{Transport Measurements}
Electrical and thermoelectric transport measurements were carried out in a pumped Helium cryostat with 9~T magnetic field. Magnetotransport measurements were performed using a bias current of $10-100$~nA using an SR830 low-frequency lock-in amplifier at $17.81$~Hz. Local Joule heating was employed for thermoelectric measurements to create a temperature gradient across the TBG/WSe$_2$ region. A range of sinusoidal currents ($300-400$ nA) at an excitation frequency of $\omega$~$=$ $17.81$~Hz was allowed to pass through the SLG region to create a temperature difference $\Delta T$ between two ends of the TBG/WSe$_2$ channel and the resulting $2^\mr{nd}$ harmonic thermo-voltage ($V_{2\omega}$) was recorded using the same lock-in amplifier. To estimate the twist angle, we used the relation, $n_s=8\theta^2/\sqrt{3}a^2$ where $a = 0.246$ nm is the lattice constant of graphene and $n_s$ ($\nu = 4$) is the charge carrier density corresponding to a fully filled superlattice unit cell. 

\subsection*{Real-space lattice calculations}
The atomic positions were relaxed using LAMMPS molecular dynamics simulations~\cite{PLIMPTON19951} based on EXX-RPA-informed~\cite{PhysRevB.96.195431} force-fields, and the electronic structure was obtained through real-space tight-binding calculations as outlined in the supplementary information. Both the band structure and DOS were calculated by a shift-invert diagonalization method~\cite{Arpack} with a broadening of $0.8$ meV, while the band structures in the absence of SOC terms are calculated using exact diagonalization. We used a 4x3x1 Monkhorst-Pack k-point grid for the DOS calculation in a rectangular supercell containing 22328 atoms. 
The graphene itself is modeled using the F2G2 model~\cite{PhysRevB.87.195450}. More details on the Hamiltonian implementation, analysis of the different terms in the Hamiltonian and phase shifted patterns can be found in the supplementary information. 

\section*{Acknowledgements}
We gratefully acknowledge the usage of the MNCF and
NNFC facilities at CeNSE, IISc. UC acknowledges funding from SERB via grants ECR/2017/001566 and SPG/2020/000164. 
NL was supported by the Korean National Research Foundation grant NRF-2020R1A2C3009142 
and AS by grant NRF-2020R1A5A1016518. 
DL was supported by the Korean Ministry of Land, Infrastructure and Transport(MOLIT) from the Innovative Talent Education Program for Smart Cities.
JJ was supported by the Samsung Science and Technology Foundation under project SSTF-BAA1802-06. We acknowledge computational support from KISTI through grant   
KSC-2021-CRE-0389 and the resources of Urban Big data and AI Institute (UBAI) at the University of Seoul. KW and TT acknowledge support from the Elemental Strategy Initiative
conducted by the MEXT, Japan (Grant Number JPMXP0112101001) and  JSPS
KAKENHI (Grant Numbers JP19H05790 and JP20H00354).
\\

\section*{Author Contributions}
SB fabricated the device, performed the measurements, and analyzed the data. BG, MP, and PSM assisted in measurements/analysis. KW and TT grew the hBN crystals. AG and UC advised on experiments. NL, SA, DL, and JJ performed the theoretical calculations. SB, NL, JJ, and UC wrote the paper with inputs from other authors. 

\section*{Competing interests}
The authors declare no competing interests.

\section*{Data availability}
The data that support the findings of this study are available from the corresponding author upon reasonable request.

\section*{Code availability}
The code that support the findings of this study are available from the corresponding author upon reasonable request.

\bibliographystyle{naturemag}
\bibliography{References}

\newpage
\clearpage

\onecolumngrid
\begin{center}
\textbf{SUPPLEMENTARY INFORMATION}
\end{center}

\maketitle
\section{Simulation}
{Here, we outline the details of the tight-binding hopping terms, in particular the interlayer coupling that we have used in our simulations. 
We then show the band structures corresponding to a shifted origin 1D potential that shows the relative robustness of band folding features. We further present band structures for different Hamiltonian model parameters without and with spin-orbit coupling (SOC) terms to better understand the role of the different terms in the Hamiltonian.

\subsection{Twisted bilayer graphene's tight-binding Hamiltonian}

The twisted bilayer graphene (TBG) interlayer hopping terms between atoms at lattice points $\textbf{R}_i$ and $\textbf{R}_j$ are given by~\cite{1910.12805}
\begin{equation}
%\begin{split}
   -t_{lm} = -t(\textbf{R}_l,\textbf{R}_m)
   %= S(\textbf{d} \cdot  \textbf{e}_z)
   =   S_{0} \exp[(\textbf{d} \cdot \textbf{e}_z-p)/q] 
   %\\ 
   \times  
   \left[V_{pp\sigma}(d) \left(\frac{ \textbf{d} \cdot \textbf{e}_z}{d} \right)^2 \right].
\label{distDependentHopping}
%\end{split}
\end{equation}
with $\textbf{d} = \textbf{R}_l - \textbf{R}_m $, $d = \left| \mathbf{d} \right|$ and $\textbf{e}_z$ the unit vector parallel to the $z$ axis 
and where~\cite{PhysRevB.86.125413}
\begin{equation}
    V_{pp\sigma}(d) = V_{pp\sigma}^0 \exp\left(-\frac{(d-d_0)}{r_0}\right)
\label{vppsigmaEq}
\end{equation}
with $d_0$ the interlayer distance, $a_0$ the interatomic carbon distance, $r_0 = 0.184 a$
%$V_{pp\pi}^0 = 2.7$ eV is the transfer integral between nearest-neighbor atoms 
and $V_{pp\sigma}^0 = -0.48$ eV is the transfer integral between two vertically aligned atoms. This form in Eq.~\ref{vppsigmaEq} is widely used in the literature~\cite{PhysRevB.85.195458} and is good at reproducing the largest magic angle in TBG, even though the corresponding effective hopping term, obtained by summating intralayer intra-sublattice terms~\cite{PhysRevB.87.195450}, of $-2.45$ eV (when using $V_{pp\pi}^0 = 2.7$ eV for the transfer integral between nearest-neighbor atoms) corresponds to a Fermi velocity which is much smaller than experiment or even local density approximation (LDA). To correct for this approximate model in square brackets, we introduced the variables $p=1.34$ and $q=3.25$ to better capture the inter-layer distance modulations that enter the fitting procedure to the DFT tunneling at the K-point for all slidings as well as the magic angle renormalization factor $S_0$ equal to $0.895$ for these structures that we relaxed using LAMMPS MD relaxations~\cite{PLIMPTON19951} using EXX-RPA-informed~\cite{PhysRevB.96.195431} force fields. This choice of $S_0$ value leads to the maximum flatness of bands at $1.08^\circ$ when all terms but the first are equal to zero in the Hamiltonian from Eq.~(2) in the manuscript. For the intralayer terms we simply use the F2G2 model~\cite{PhysRevB.87.195450}.

\subsection{Doubling of the bands due to band-folding}
In Fig.~S1, we illustrate how the existence of a potential that doubles the initial TBG moir\'{e} period by a factor two effectively leads to band-folding and associated doubling of the initial 4-fold degenerate valence and conduction bands. We calculate the band structures in the smallest moire triangular unit cell and compare with the folded bands with twice the original unit cell area where we can observe how the initial bands are being doubled in number. We neglect any spin/valley degeneracy lifting terms in the Hamiltonian and will discuss their effect further in Sect.~\ref{HamTerms}.

\begin{figure*}[bth]
\includegraphics[width=0.4\textwidth]{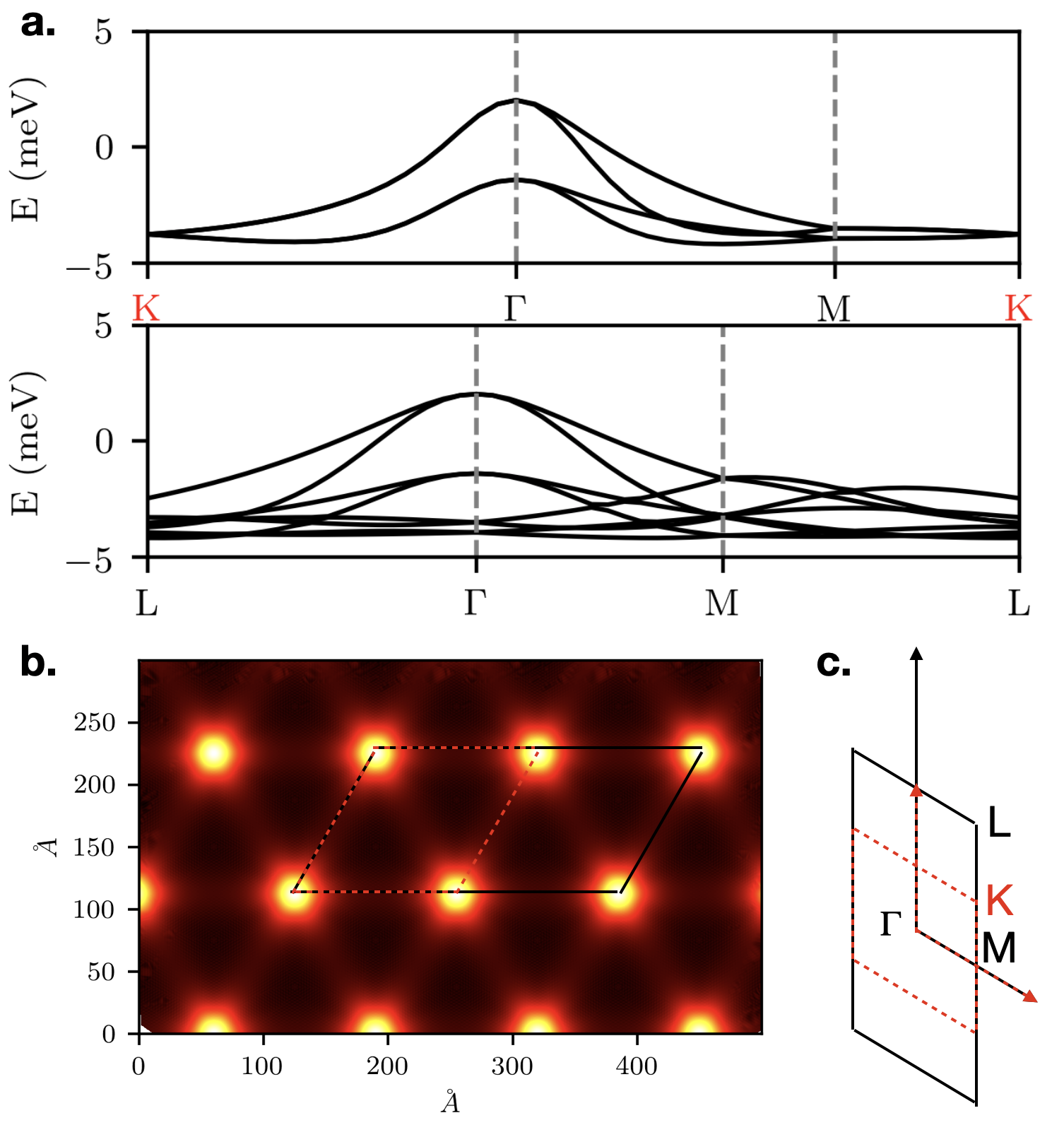}
%\captionsetup{justification=raggedright,singlelinecheck=false}
\justify{\textbf{Supplementary Fig.} S1. \textbf{Band folding and doubling of bands}. 
\textbf{a.} The top panel illustrates the bandstructure for the unperturbed original triangular moir\'{e} cell illustrated by the parallelogram linking 4 closest AA-stacking bright dots in \textbf{b.}, whereas the bottom panel shows the band structure for the doubled area supercell. The spin is neglected in this figure, as such we see a total of 4 bands (two valleys for the conduction band and valence band) in the top panel becoming 8 bands by band-folding in the bottom panel. \textbf{b.} Real-space schematic of the single and double area supercells indicated by the dashed red and black solid parallelograms. {\textbf c.} The associated BZ for the band structures for the single and double area supercells indicated also with dashed red and black solid lines.}
\end{figure*}

\subsection{Alternative 1D potential modulation}
We have chosen a specific form of the 1D sine potentials in the main text to obtain the folded bands in the larger period cell. Here, we provide an additional 1D potential that leads to a doubled supercell area compared to the moir\'{e} pattern unit cell area that leads to quantitatively different types of folded bands. 
In Fig.~S2, we apply a cosine instead of a sine 1D modulation in Eq.~2 in the manuscript (set $\phi = \pi/2$). In this case, the AA stacking regions are located where the 1D potential is maximal or minimal.

\begin{figure*}[bth]
\includegraphics[width=0.6\textwidth]{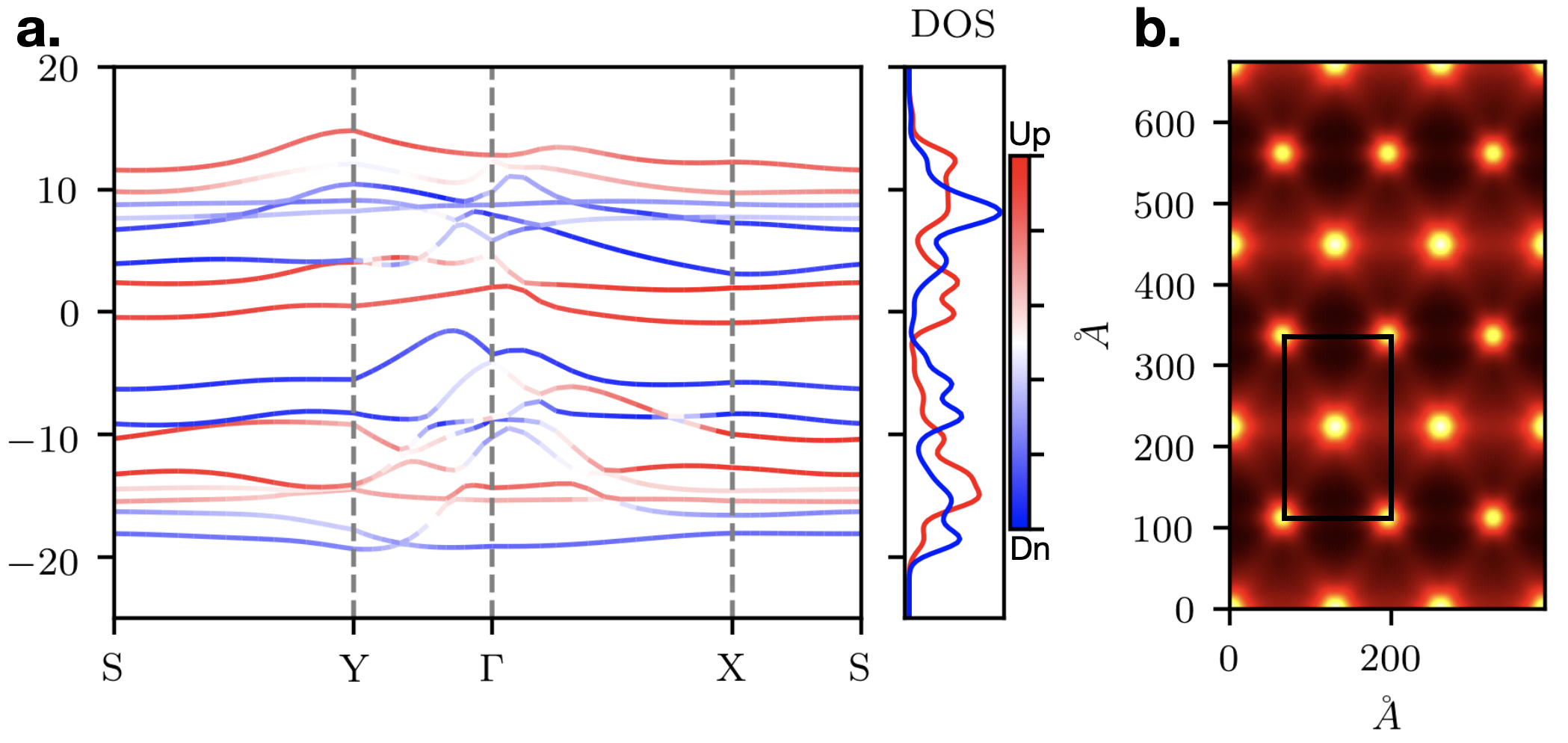}
%\captionsetup{justification=raggedright,singlelinecheck=false}
\justify{\textbf{Supplementary Fig.} S2. \textbf{Cosine modulation for 1D potential in the rectangular unit cell}. \textbf{a.} Bandstructure of TBG under a 1D cosine periodic potential in Eq.~2 by setting $\phi = \pi/2$, and using the same parameters $C_z = 30$ meV, $t_2 = 15$ meV, $\lambda_R = 10.5$ meV and $\Delta = 300$~meV. \textbf{b.} Colormap illustrating the cosine modulation in the rectangular unit cell using the same procedure as in the main text. The high-symmetry path in the Brillouin zone (BZ) is the same as in the manuscript.} 
\end{figure*}

\begin{figure*}[bth]
\includegraphics[width=0.7\textwidth]{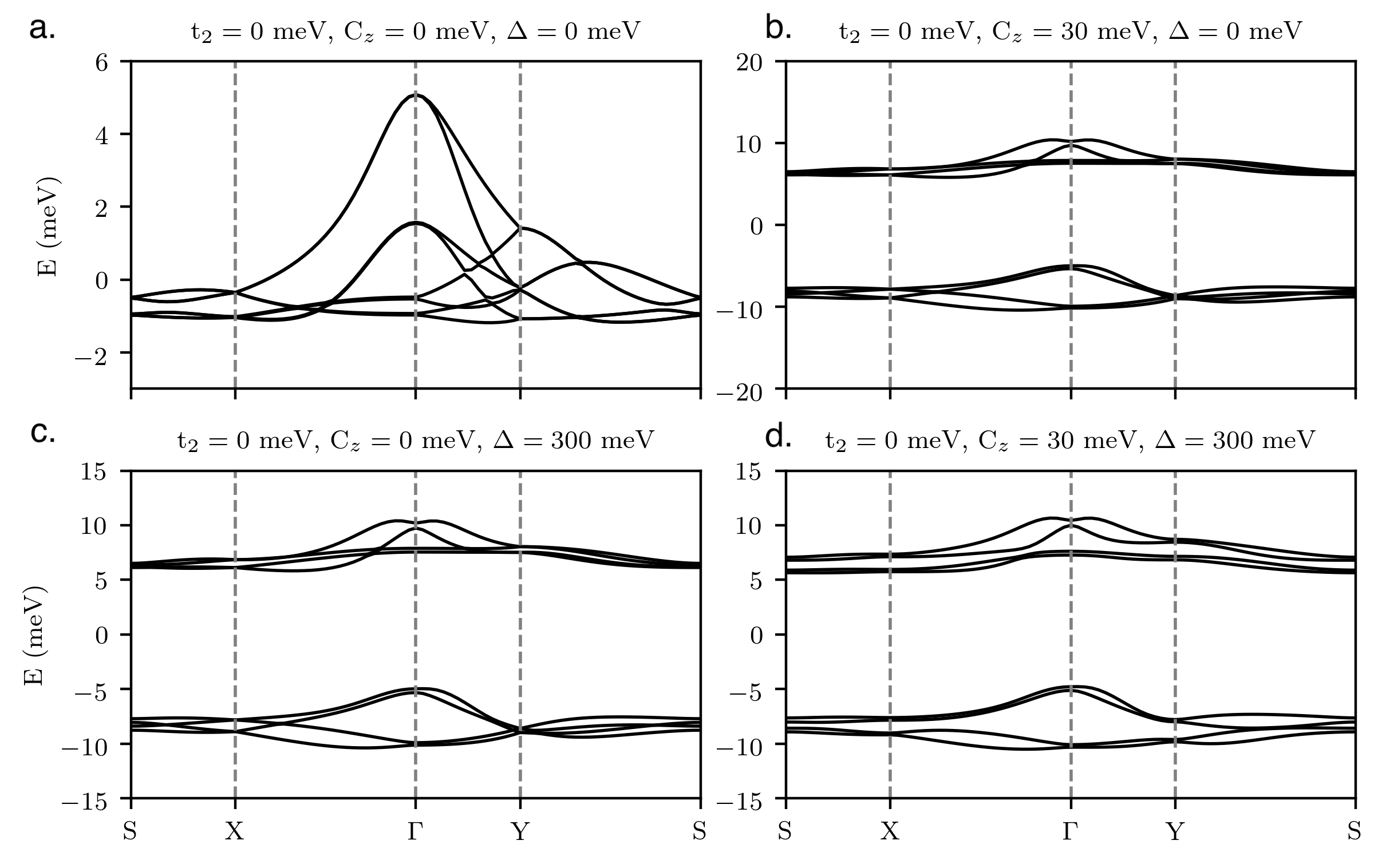}
%\captionsetup{justification=raggedright,singlelinecheck=false}
\justify{\textbf{Supplementary Fig.} S3. \textbf{Impact of the different Hamiltonian terms}. We gradually include the different terms from the Hamiltonian in Eq.~2 from the manuscript, while ignoring the Kane-Mele term by setting $t_2 = 0$ as well as the Rashba term by setting $\lambda_R=0$ meV. Comparison of these figures with the final band structure in the manuscript illustrates that each term of this heuristic Hamiltonian can play an important role.}
\end{figure*}

\begin{figure*}[bth]
\includegraphics[width=1\textwidth]{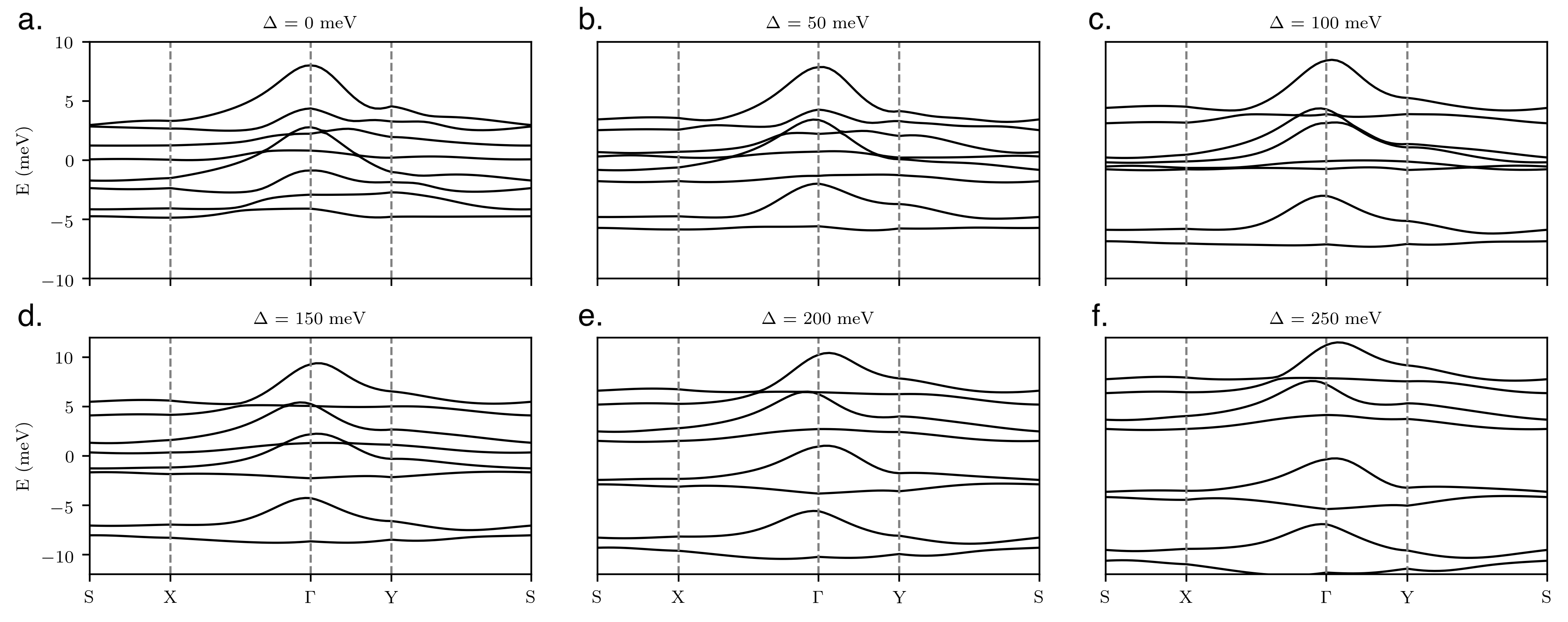}
%\captionsetup{justification=raggedright,singlelinecheck=false}
\justify{\textbf{Supplementary Fig.} S4. \textbf{Effect of the constant mass-term in the band separation}. We gradually increase the constant mass-term, with $t_2 = 3$ meV and $\lambda_R=0$~meV, we only show the spin up contribution here, and $C_z = 0.03$~meV in Eq.~2 from manuscript. The constant mass term clearly increases the separation between the conduction and valence band groups that is expected to reduce interband screening and therefore enhance the chances of opening a band gap due to Coulomb interactions when the subbands are completely filled.}
\end{figure*}

%\newpage
\subsection{Role of the different Hamiltonian terms}
\label{HamTerms}

In Figs.~S3, S4, and S5, we revisit the band-structure shown in the manuscript main text consisting of a rectangular unit cell and 
sine 1D modulation potential to illustrate the effect of the different terms in our Hamiltonian in more detail.
In Fig.~S3, we divide in different panels the band structures and start by showing in Fig.~S3a 
the band structure for the TBG, nearly flat bands in the rectangular unit cell. 
The area of this rectangular unit cell is twice as large as the area of the single TBG system (smallest parallelogram delimited by 4 AA-stacking regions); hence we observe a doubling in the number of the flat bands. In the absence of spin, but including the valley degeneracy (inherent to our real-space approach), we have 8 flat bands in the system, distributed evenly among conduction and valence bands. In Fig.~S3b, we include the 1D sine modulation pattern and observe a clear energy separation of the conduction and valence bands. Alternatively, in Fig.~S3c, we check the effect of adding a constant mass-term in the bottom graphene layer using the rectangular unit cell in the absence of a sine modulation. We see that in this case, the conduction bands and valence bands are being separated from each other too, but for our choice for the value of $\Delta$, the doubled conduction bands remain very close to each other and cannot be disentangled at the meV energy level (same observation for the doubled valence bands). We finally combine in Fig.~S3d the effect of both the sine potential and the constant mass-term. The 4 valence and 4 conduction bands get separated into 2 sets of 2 bands each. We note that the weak band degeneracy lifting at the high symmetry points originates in the $\xi_{c_i}$ factor, which differentiates between sublattices that we have included in our 1D sine modulation. We have checked that in the absence of this factor, this degeneracy lifting is nonexistent. Finally, for the case in Fig.~S3d, only the Rashba and the Kane-Mele term are missing when compared to the band structure from the paper; this shows that for this choice of parameters, these terms are crucial in their ability to resolve the degeneracy lifted spins that are causing the features at $\nu=0.5$ and $3.5$. In Fig.~S4, we further discuss qualitatively the effect of either the mass-term in the presence of a constant Kane-Mele term, and 
in Fig.~S5, we vary the magnitude of the Kane-Mele term that can arise due to proximity effects induced by the substrate. We omit the Rashba term for simplicity here and only show the spin-up component. We notice that while the mass-term tends to push the highest set of valence bands above the lowest set of the conduction band, the Kane-Mele term shifts them in the opposite directions. The combination of these two terms recovers the original ordering of the bands, and both terms tend to widen the energy spectrum range where 
the nearly flat bands distribute.

\begin{figure*}[bth]
\includegraphics[width=0.6\textwidth]{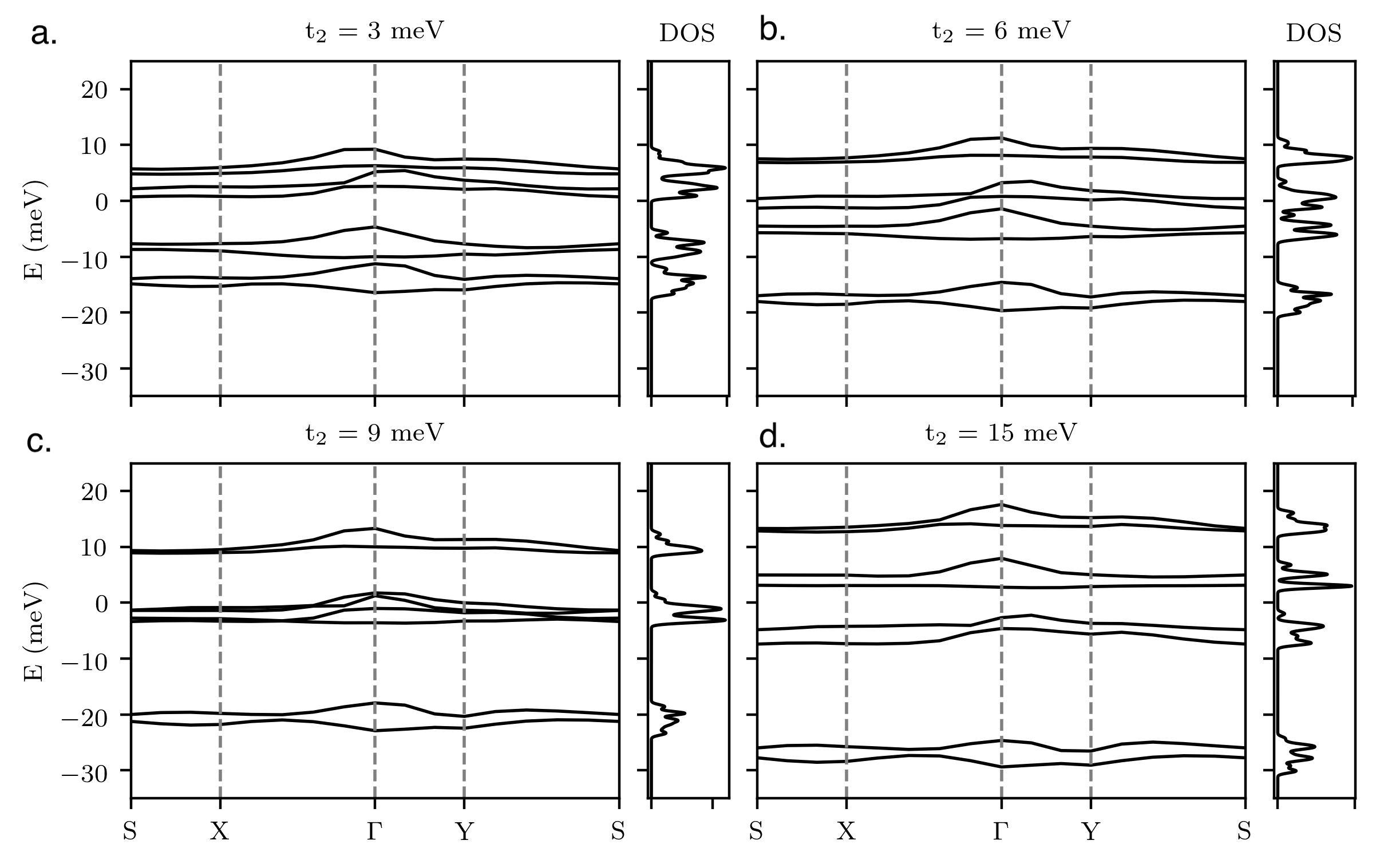}
%\captionsetup{justification=raggedright,singlelinecheck=false}
\justify{\textbf{Supplementary Fig.}~S5. \textbf{Effect of the the value of $t_2$ controlling the strength of the Kane-Mele term.} We gradually increase the value of the Haldane pre-factor $t_2$ entering the Kane-Mele term, with $C_z = 30$~eV, $\Delta = 300$~meV and $\lambda_R = 0$~eV, in Eq.~2 from manuscript (we only show the spin-up component). Increasing the value of $t_2$ resolves the degeneracy of the spin-up and spin-down bands in momentum space although it does not lead to an overall splitting in the DOS.}
\end{figure*}
}
\newpage
\section{Experimental Results}

{\begin{figure*}[bth]
\includegraphics[width=1.0\textwidth]{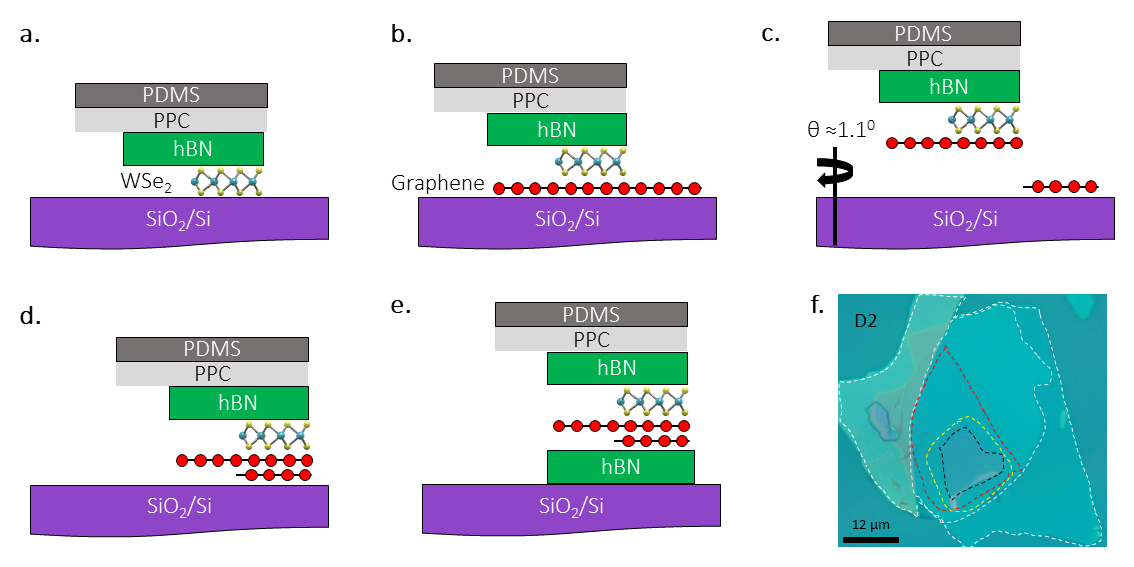}
%\captionsetup{justification=raggedright,singlelinecheck=false}
\justify{\textbf{Supplementary Fig.}~S6. \textbf{Device fabrication}. \textbf{a-e.} Schematic of the various steps in the ‘tear and stack’ method used for fabrication of the moir\'{e} heterostructure. A layer of PDMS/PPC on a glass slide is used to sequentially pick the various heterostructure layers. A twist angle of $\approx 1.1^{\circ}$ is used to create the magic-angle twisted bilayer graphene. \textbf{f.} Optical image of the final stack for device D2. The top and bottom graphene (red and yellow), WSe$_2$ (black) and two hBN (white) layers are shown using dashed lines. The scale bar is $12$~$\mu$m.}
\end{figure*}
%\noindent

\begin{figure*}[bth]
\includegraphics[width=1.0\textwidth]{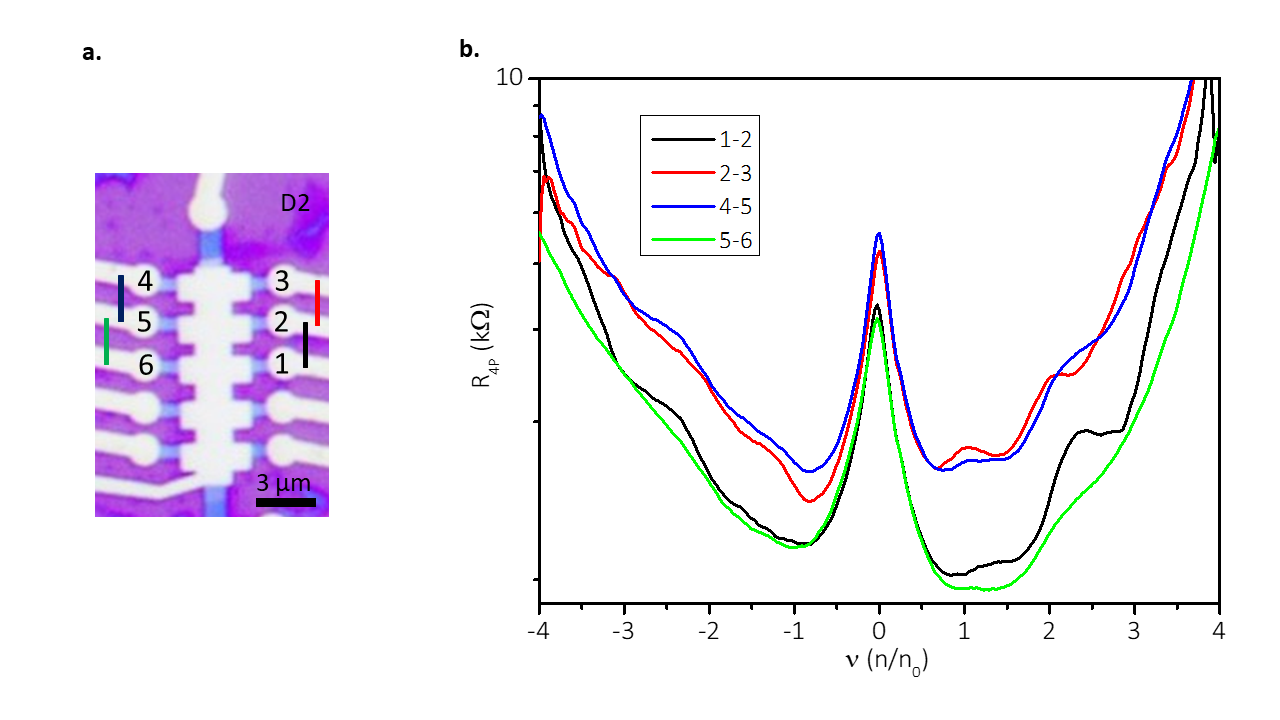}
%\captionsetup{justification=raggedright,singlelinecheck=false}
\justify{\textbf{Supplementary Fig.}~S7. \textbf{Characterization of device D2}. \textbf{a.} Optical image of device D2 where the scalebar denotes 3 $\mu$m. \textbf{b.} Four-probe resistance measurements at $T =$ 6 K for different pair of contacts show twist angle uniformity throughout the TBG/WSe$_2$ channel. Different colors have been used for different pairs of voltage probes. Cr/Au top gate (see (a)) was used to tune the carrier density in the Hallbar.}
\end{figure*}

\begin{figure*}
\includegraphics[width=1.0\textwidth]{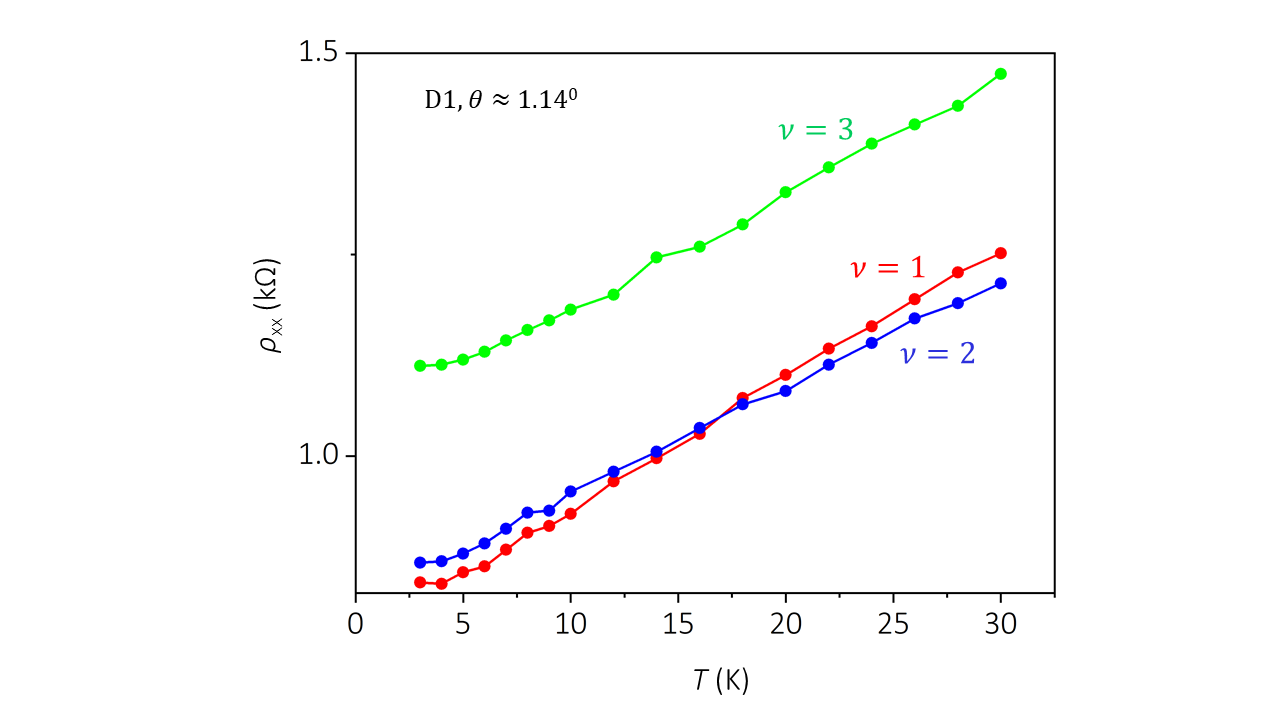}
%\captionsetup{justification=raggedright,singlelinecheck=false}
\justify{\textbf{Supplementary Fig.} S8. \textbf{Linear-in-$T$ resistance in device D1, $\theta \approx 1.14^\circ$} The longitudinal resistivity $\rho_{xx}$ at various filling factors $\nu=1, 2$ and 3 vary linearly with $T$ in the measured range of temperature (3 - 32 K). The slopes for $\nu=1$, 2, and 3 were found to be 16, 13, and 14 $\Omega/K$ respectively.}
\end{figure*}

\begin{figure*}
\includegraphics[width=1.0\textwidth]{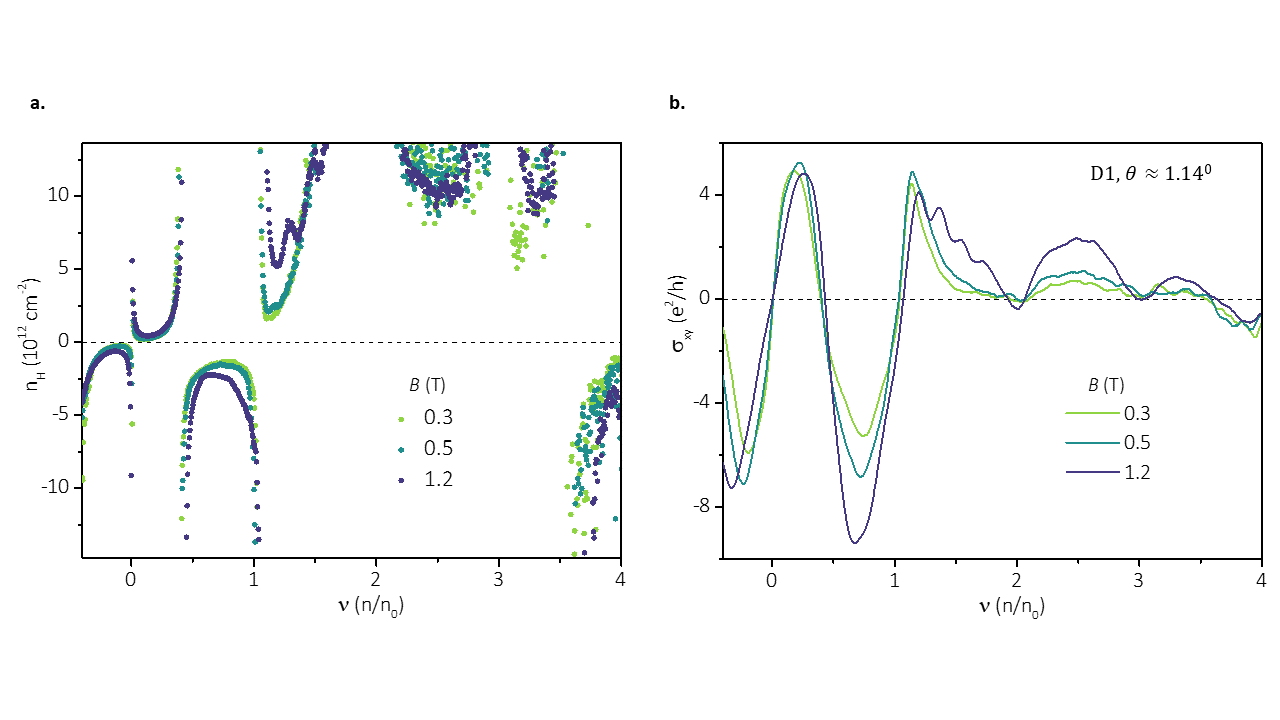}
%\captionsetup{justification=raggedright,singlelinecheck=false}
\justify{\textbf{Supplementary Fig.} S9. \textbf{Low-field Hall data in device D1, $\theta \approx 1.14^\circ$}. \textbf{a.} $n_H-\nu$ plot for $B = 0.3 - 1.2$ T. \textbf{b.} $\sigma_{xy}$ as a function of $\nu$ for the same range of $B$. The zero crossings at $\nu = 0.5,1$ and 3.5 and reset at $\nu = 2$ and 3 are robust with $B$.}
\end{figure*}

\begin{figure*}
\includegraphics[width=1.0\textwidth]{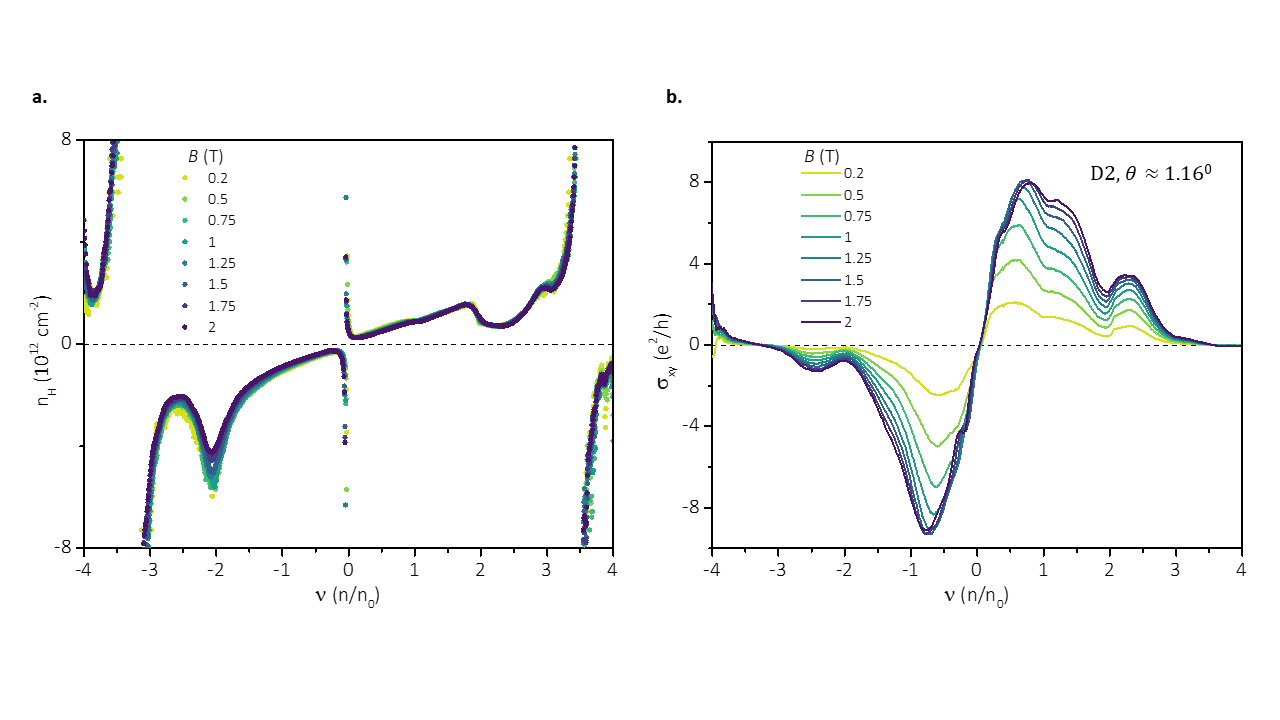}
%\captionsetup{justification=raggedright,singlelinecheck=false}
\justify{\textbf{Supplementary Fig.} S10. \textbf{Low field Hall data in device D2, $\theta \approx 1.16^\circ$}. \textbf{a.} $n_H-\nu$ plot for $B = 0.2 - 2$ T. The sign-change at $\nu = 3.5$ appears at the lowest $B$-field of 0.2 T. In addition, the reset at charge carriers is observed at $\nu = \pm2$. \textbf{b.} $\sigma_{xy}$ as a function of $\nu$ for the same range of $B$.}
\end{figure*}

\begin{figure*}
\includegraphics[width=1.0\textwidth]{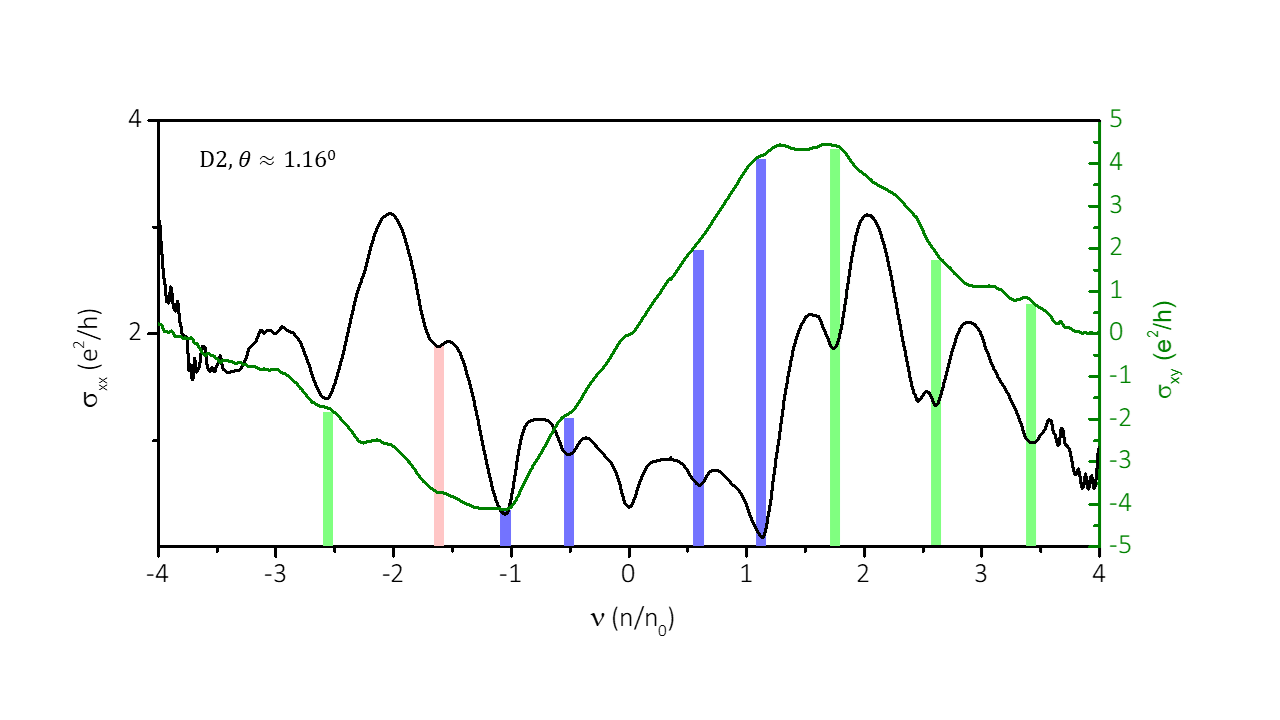}
%\captionsetup{justification=raggedright,singlelinecheck=false}
\justify{\textbf{Supplementary Fig.} S11. \textbf{Hall conductivity at $B = 9$ T in device D2}. \textbf{a.} $\sigma_{xy}$ shows plateaus as $\sigma_{xy} = Ce^2/h$ associated with the minima in $\sigma_{xx}$. Different color bars (blue for the CNP, green for $\nu = 1, \pm2, 3$ and red for $\nu = -0.5$) have been used to show the sequence of symmetry-broken quantized states nucleating from several partial fillings of flat bands.}
\end{figure*}

\begin{figure*}
\includegraphics[width=0.8\textwidth]{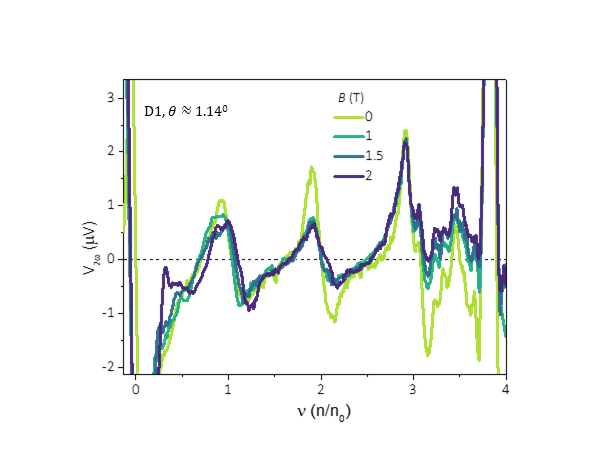}
%\captionsetup{justification=raggedright,singlelinecheck=false}
\justify{\textbf{Supplementary Fig.} S12. \textbf{Magneto-thermoelectricity measurements in device D1}. Thermoelectric voltage $V_{2\omega}$ measured at $B = 0, 1, 1.5$, and 2~T and $T = 3$ K for a heating current of 300 nA. The feature at $\nu=3.5$ is clearly visible at all $B$.}
\end{figure*}
}

\end{document}